\documentclass{IEEEtran}
\ifCLASSINFOpdf
\else
\fi
\usepackage{multicol}
\usepackage{gensymb}
\usepackage{lipsum}
\usepackage{xcolor}
\usepackage{multirow}
\usepackage{enumitem}
\usepackage{geometry}
\usepackage{graphicx}
\usepackage{subcaption}
\usepackage{adjustbox}
\usepackage{graphicx}
\usepackage{float} 
\usepackage{array, booktabs, multirow}

\usepackage{unicode-math}
\usepackage{threeparttable}
\makeatletter
\def\footnoterule{\kern-3\p@
	\hrule \@width 0.49\textwidth \kern 2.6\p@}
\makeatother
\usepackage{amsmath}
\usepackage{caption}
\usepackage{makecell}
\newcommand{\subparagraph}{}
\usepackage{titlesec}
\titlespacing{\section}
{0pt}{5pt}{5pt}
\titlespacing{\subsection}
{0pt}{5pt}{5pt}

\begin{document}

\begin{flushleft}
	\fontsize{14}{15} \fontfamily{phv}\selectfont
	 \textbf{A 24-GHz CMOS Transformer-Based Three-Tline Series Doherty Power Amplifier Achieving 39\% PAE}
\end{flushleft}

\begin{flushleft}	
	{\fontsize{10}{10} \fontfamily{phv}\selectfont 
		Zheng Wang\textsuperscript{1}, Yifu Li\textsuperscript{1}, Yuchao Mei\textsuperscript{1}, Xinyu Sui\textsuperscript{1}, Qingbin Li\textsuperscript{1}, Xu Luo\textsuperscript{1}, Rui Wang\textsuperscript{1}, Dongxin Ni\textsuperscript{2}, Jian Pang\textsuperscript{1}\\
        \textsuperscript{1}State Key Laboratory of Radio Frequency Heterogeneous Integration, Shanghai Jiao Tong University\\
		\textsuperscript{2}School of Microelectronics, Nanjing University of Science \& Technology
	}
\end{flushleft}
	\noindent
	\begin{abstract}
        This paper presents a transformer-based three-transmission-line (Tline) series Doherty power amplifier (PA) implemented in 65-nm CMOS, targeting broadband K/Ka-band applications. By integrating an impedance-scaling network into the output matching structure, the design enables effective load modulation and reduced impedance transformation ratio (ITR) at power back-off when employing stacked cascode transistors. The PA demonstrates a -3-dB small-signal gain bandwidth from 22 to 32.5 GHz, a saturated output power (Psat) of 21.6 dBm, and a peak power-added efficiency (PAE) of 39\%. At 6dB back-off, the PAE remains above 24\%, validating its suitability for high-efficiency mm-wave phased-array transmitters in next-generation wireless systems.
	\end{abstract}
	
	\noindent

\section{Introduction}

The explosive growth in wireless communication technologies, including 5G and satellite communication, has motivated the deployment of high-data-rate mm-wave front ends in the K(18–26.5 GHz) and Ka-band (26.5–40 GHz). These systems typically require power amplifiers (PAs) capable of supporting modulations with large peak-to-average power ratios (PAPRs), such as 64-QAM and 256-QAM, while maintaining energy efficiency and linearity under significant power back-off (PBO).\par

Among various linearity and efficiency enhancement techniques, the Doherty Power Amplifier (DPA) stands out as a preferred architecture in mm-Wave bands due to its ability to enhance back-off efficiency without requiring additional supply rails or high-speed modulators. Meanwhile, DPAs often adopt cascode topology that boosts gain [3,6], enables higher voltage swings, and increases saturated output power. \par

\section{Proposed Transformer-based Three-Tline Teries Doherty PA}

Traditional DPAs use parallel $\lambda/4$-line-based combiners which are area-inefficient and suffer from high ITR under back-off. As shown in Fig. 1, the previous three-line parallel DPA [1] addresses the issue of high ITR under PBO conditions, but the parallel topology inherently requires the PA to exhibit high output impedance (approaching the characteristics of an ideal current source), which poses significant challenges for implementation in the mm-wave frequency band. In contrast, the conventional series Doherty PA, while maintaining a low ITR under PBO (similar to the three-line parallel Doherty configuration), relies on voltage combining. This principle demands that the PA’s output impedance remains sufficiently low. Unfortunately, the adoption of cascode technology exacerbates the design complexity of the output matching network and narrows the operating bandwidth.\par
\begin{figure}[H]
    \centering
    \includegraphics[width=1\linewidth]{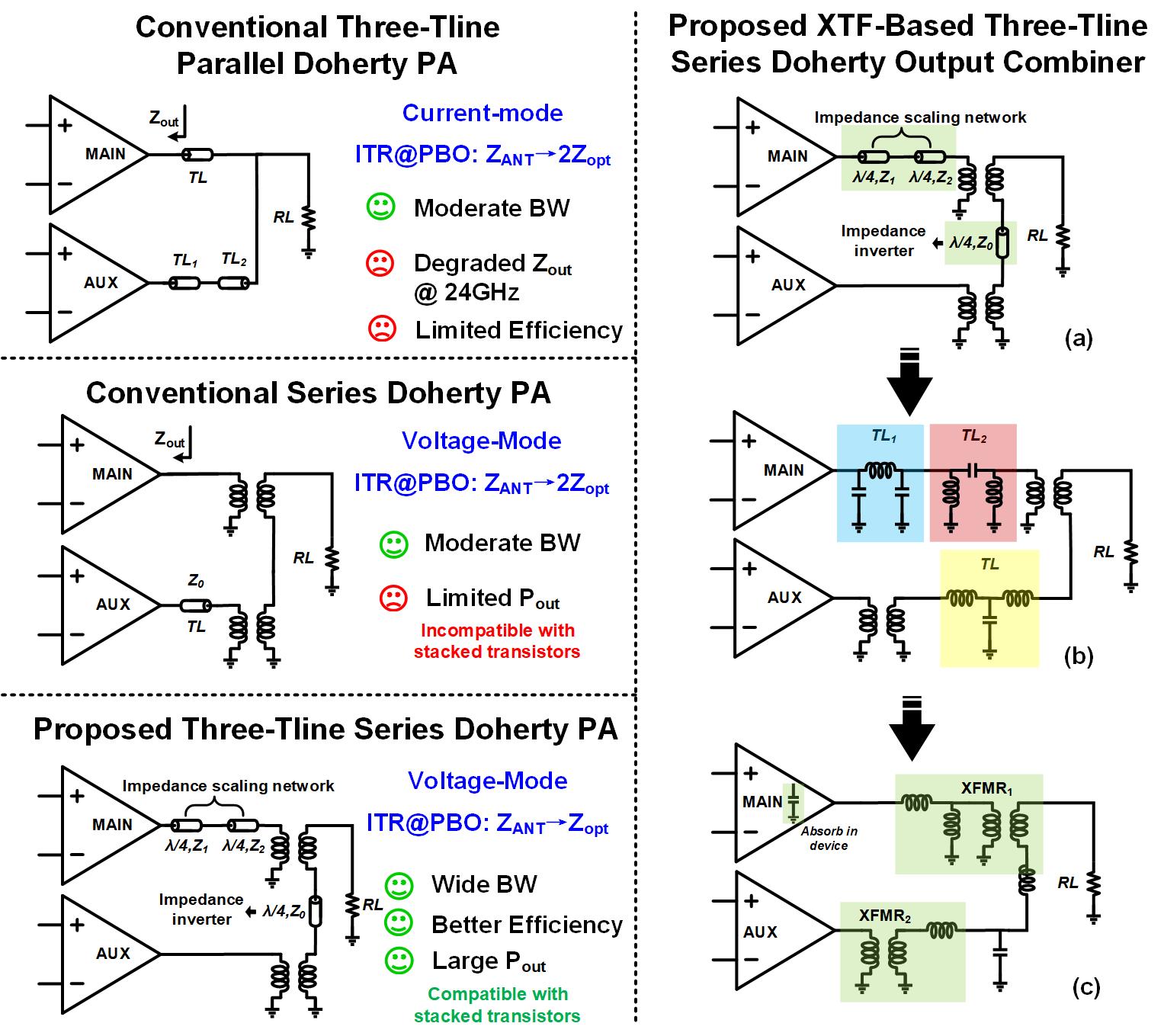}
    \caption{Concept and design procedure of the proposed transformer-based three-Tline series Doherty Output Network.}
    \label{fig:enter-label}
\end{figure}
\vspace{-0.6cm}
\begin{figure}[H]
    \centering
    \includegraphics[width=1\linewidth]{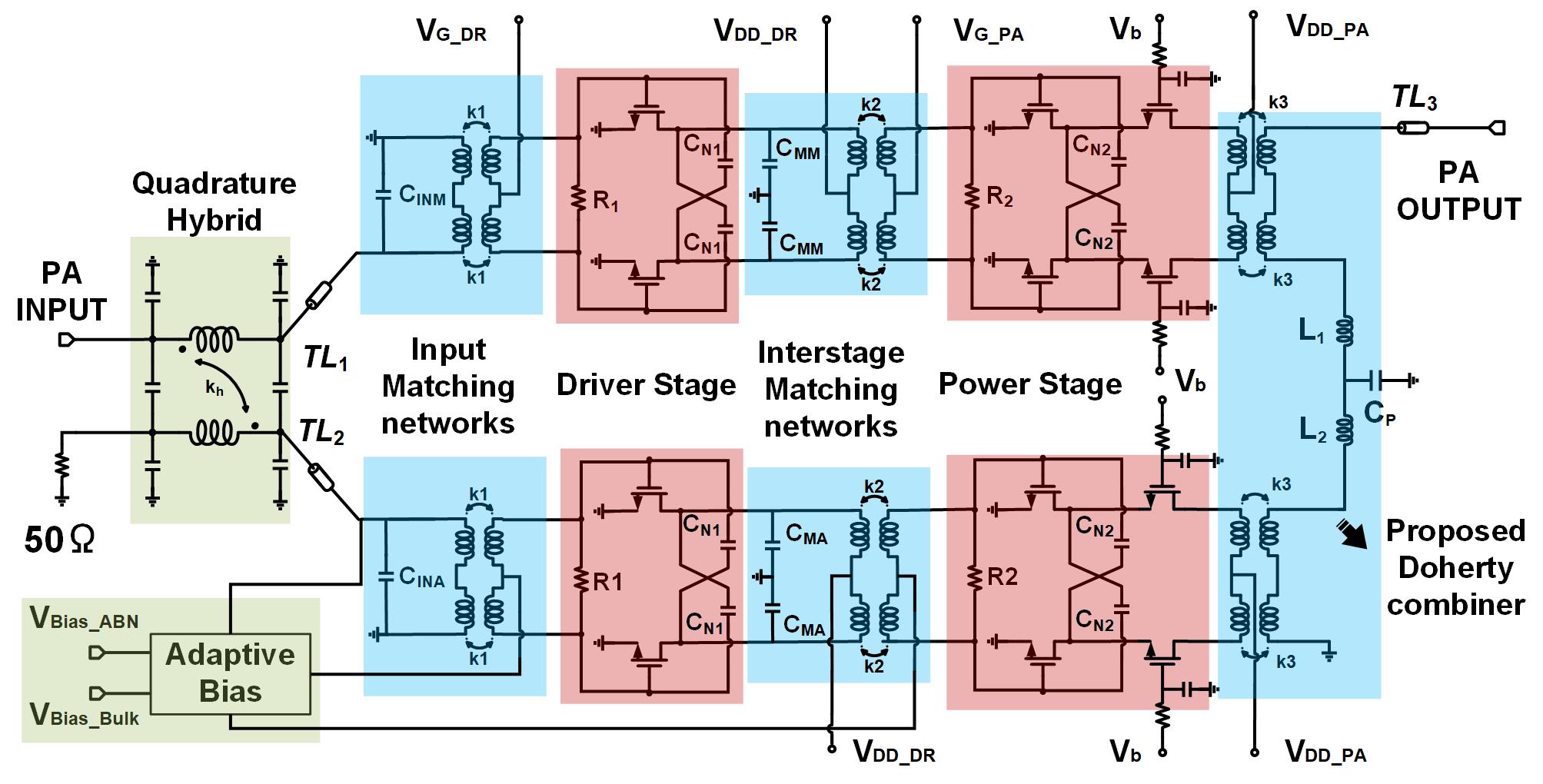}
    \caption{Schematic of the proposed transformer-based three-Tline series Doherty PA.}
    \label{fig:enter-label}
\end{figure}
\vspace{-0.6cm}
\begin{figure}[H]
    \centering
    \includegraphics[width=0.45\linewidth]{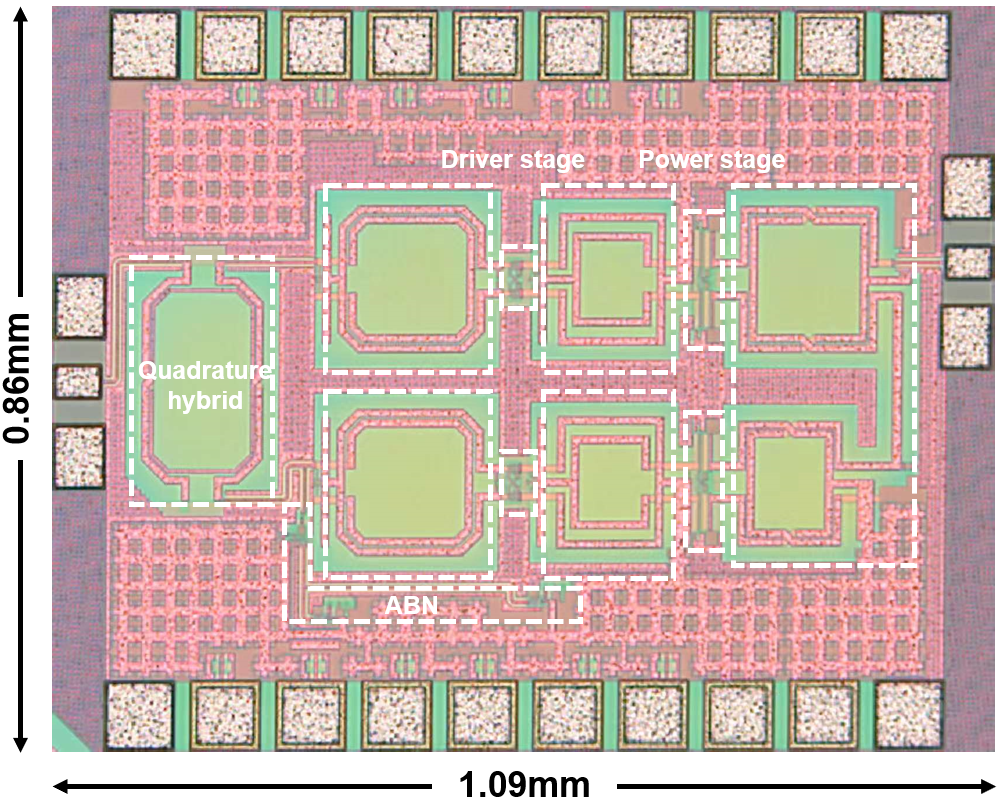}
    \caption{Chip microphotograph.}
    \label{fig:enter-label}
\end{figure}
\vspace{-0.3cm}
To address the trade-offs among bandwidth, ITR, and output power in traditional Doherty designs, this work proposes a transformer-based three-Tline series Doherty output network. By introducing an impedance scaling network into the output matching network of the conventional series Doherty architecture, the required load impedance for the main PA during the back-off region (when the auxiliary path is off) is reduced from 2Z{\text{opt}} to Zopt. Since the optimal output impedance of cascode-based PAs is typically close to the antenna port impedance (50 $\Omega$), this transformation significantly simplifies the output network design and improves broadband performance.\par

As shown in Fig. 1, the impedance inverter is realized as a T-equivalent L–C–L network, where the two inductors are absorbed into the power-combining traces of the output matching network to save area. The two $\lambda/4$-lines in the impedance scaling network are modeled using $\pi$-equivalents. The first Tline (close to the core) adopts a C–L–C configuration, where the left-side capacitor is absorbed into the parasitic capacitance of the active devices. The second Tline employs an L–C–L configuration. During design, the left-side parallel inductor of the second Tline neutralizes the right-side parallel capacitor of the first Tline; the series capacitor of the second Tline and the series inductor of the first Tline are combined to form the leakage inductance of a practical transformer; and the right-side parallel inductor of the second line becomes the transformer's magnetizing inductance. As a result, the complex impedance scaling and output matching networks are unified into a compact transformer structure, as shown in Fig. 2.

\begin{figure}[t]
    \centering
    \includegraphics[width=0.6\linewidth]{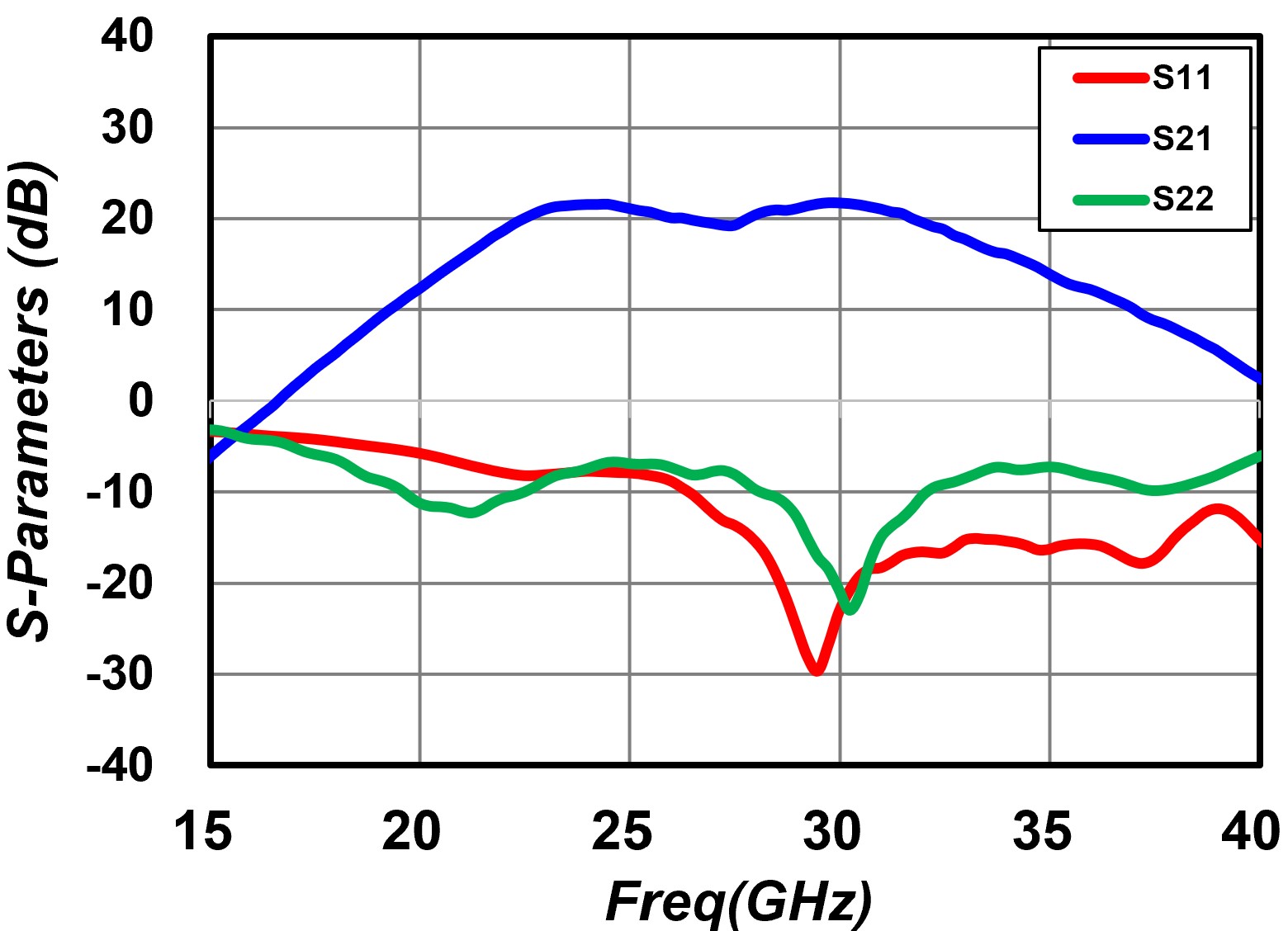}
    \caption{Measured small-signal S-parameters.}
    \label{fig:enter-label}
\end{figure}

\begin{figure}[t]
	\centering
	\begin{minipage}{0.49\linewidth}
		\centering
		\includegraphics[width=1\linewidth]{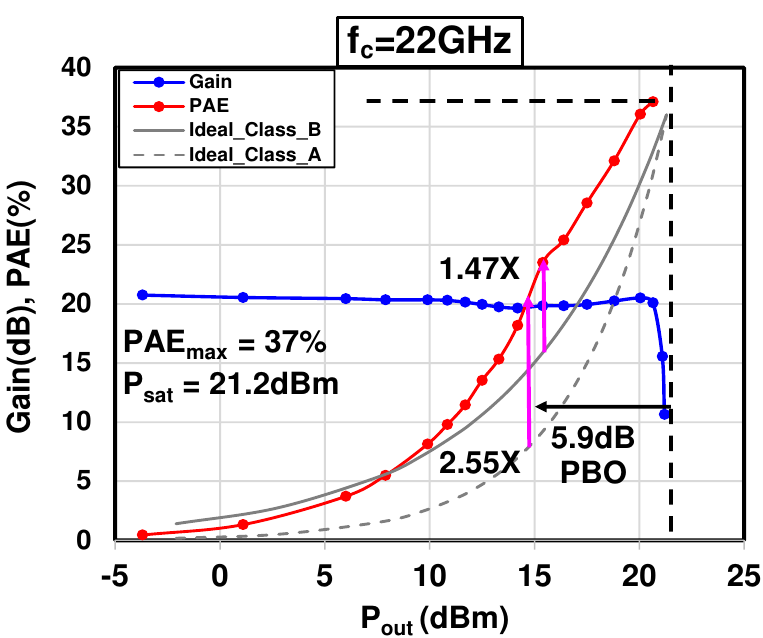}
		\parbox{0.9\linewidth}{\centering (a)} 
		\label{1a}
	\end{minipage}
	\begin{minipage}{0.49\linewidth}
		\centering
		\includegraphics[width=1\linewidth]{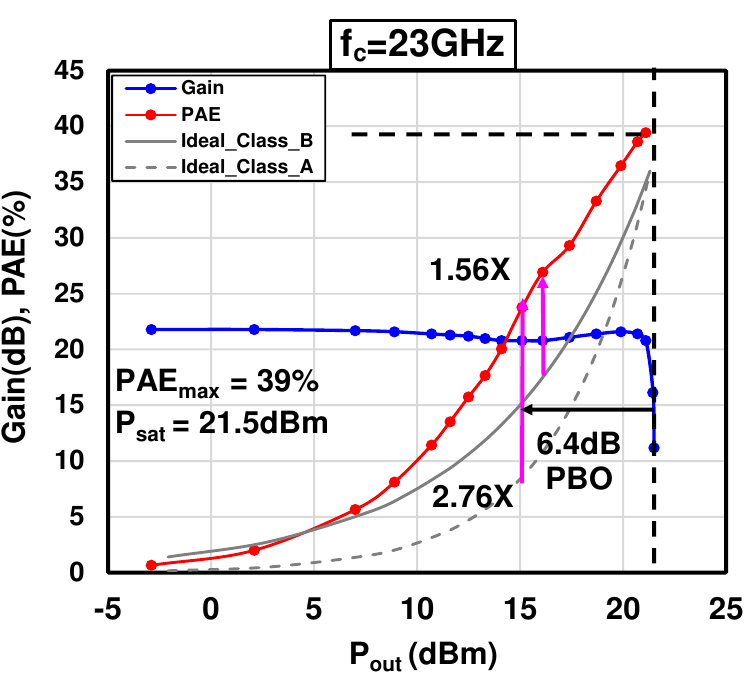}
		\parbox{0.9\linewidth}{\centering (b)} 
		\label{1b}
	\end{minipage}
    \vspace{0.1cm} 
	\begin{minipage}{0.49\linewidth}
		\centering
		\includegraphics[width=1\linewidth]{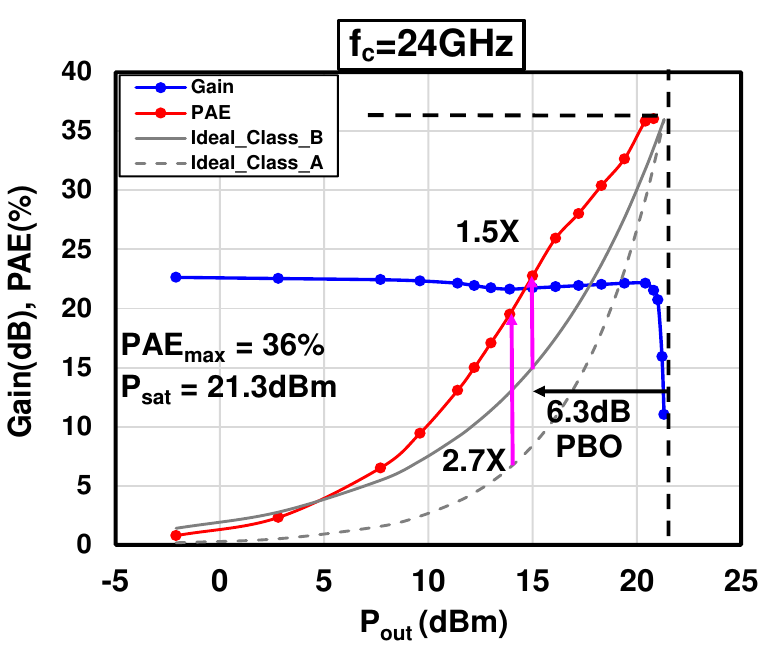} 
		\parbox{0.9\linewidth}{\centering (c)} 
		\label{1d}
	\end{minipage}
	\begin{minipage}{0.49\linewidth}
		\centering
		\includegraphics[width=1\linewidth]{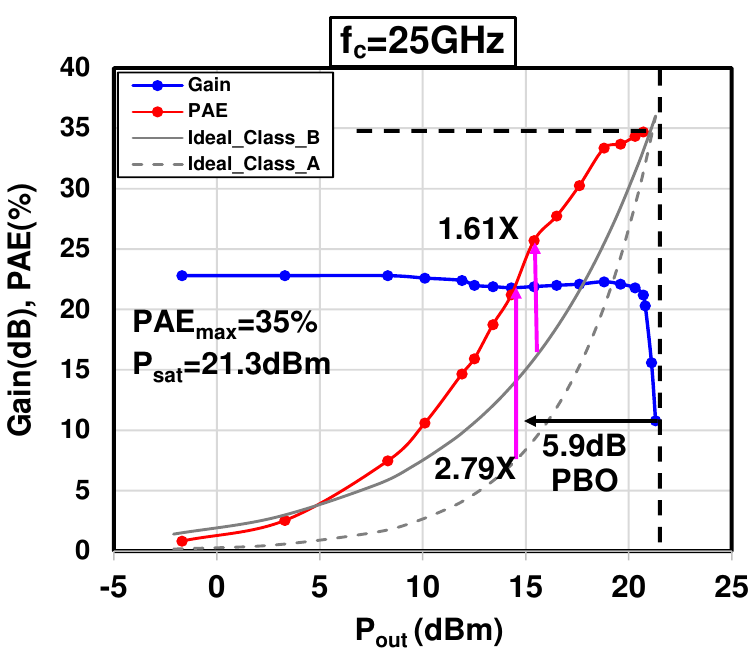}
		\parbox{0.9\linewidth}{\centering (d)} 
		\label{1e}
	\end{minipage}
    \caption{Measured large-signal performance of the Doherty PA without adjusting the ABN bias voltage from 22 to 25 GHz.}
    \label{fig1}
\end{figure}

\vspace{-0.08cm}
\section{MEASUREMENT RESULTS}
\vspace{-0.08cm}
The proposed Doherty PA is fabricated in TSMC’s 65nm process. The total area of the chip, including pads, is 0.94 mm2. Fig. 3 shows the chip photograph. The measurement results are presented as follows. Performance is summarized and compared to other state-of the-art PAs for 5G FR2 bands in Table I.
\vspace{-0.08cm}
\subsection{CW Measurement}
\vspace{-0.08cm}
Figs. 4 shows the measured small-signal S-parameters. The small-signal S21 achieves a -3-dB bandwidth of 22-32.5 GHz. Fig. 5 shows the measured PBO performance. At 23 GHz, this PA achieves 21.5 dBm Psat, 21.1 dBm P1dB, 39.4\% peak power-added efficiency and 23.7\% PAE at 6.4 dB PBO. Compared with a normalized class-B/class-A PA, the Doherty operation achieves 1.56/2.76 times PAE enhancement at 6 dB PBO. Fig. 5 also shows the measured large signal CW performance at 22, 24 and 25 GHz, respectively. Thanks to the introduced Doherty power combiner, without changing the ABN bias, significant PAE enhancements are achieved at these frequencies. The -3-dB Psat bandwidth covers 18.5-30 GHz.
\vspace{-0.08cm}
\subsection{Modulation Signal Measurements}
\vspace{-0.08cm}
Without applying DPD, the modulation signal measurement is performed with a single-carrier 64-QAM signal at 400 MSym/s symbol rates. The Doherty PA achieves -26 dB rms error vector magnitude (EVM) with average Pout/PAE of 13 dBm/17\%, as shown in Fig. 6.

\begin{figure}[t]
    \centering
    \includegraphics[width=0.45\linewidth]{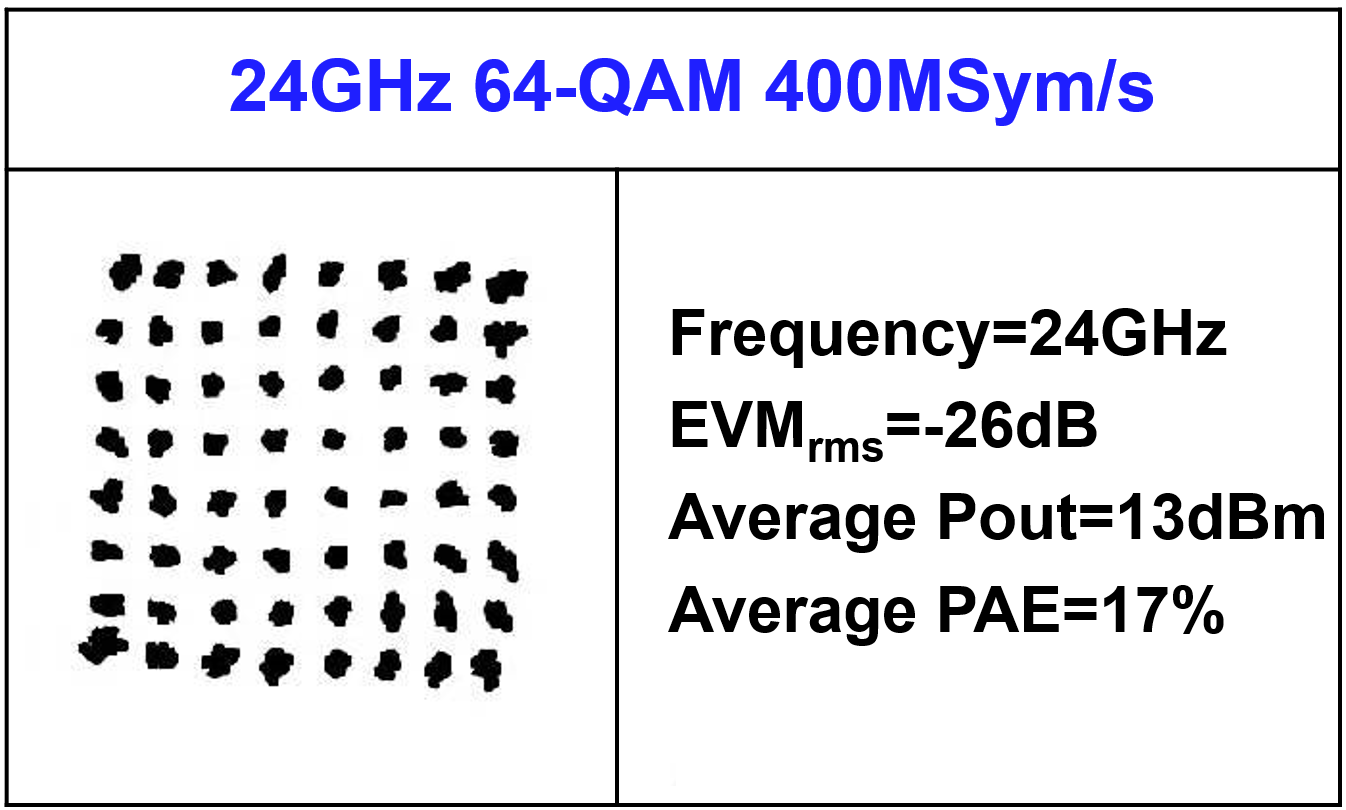}
    \caption{Measured modulation signal performance of the Doherty PA with 64-QAM modulation at a symbol rate of 400 MSym/s.}
    \label{fig:enter-label}
\end{figure}

\begin{table}[t]
\caption{Performance Comparison of mm-Wave PAs}
\centering
\resizebox{1\linewidth}{!}{
\renewcommand{\arraystretch}{1.3} 
\setlength{\tabcolsep}{3pt} 
\begin{threeparttable}
\begin{tabular}{|>{\centering\arraybackslash}m{2.3cm}|>{\centering\arraybackslash}m{2cm}|>{\centering\arraybackslash}m{2cm}|>{\centering\arraybackslash}m{2cm}|>{\centering\arraybackslash}m{2cm}|>{\centering\arraybackslash}m{2cm}|}
\hline
\rule{0pt}{0.5cm} &  \multicolumn{4}{c|}{\textbf{Mm-wave Doherty PA}} & \makecell[c]{\textbf{Mm-wave} \\ \textbf{Balanced PA} } \\ \cline{2-6} 
\rule{0pt}{0.55cm}& \textbf{This work} & \makecell[c]{[2] Chen \\TCAS I'24 }&\makecell[c]{[3] Park \\JSSC'22 } &\makecell[c]{[4] Kumaran \\TMTT'24 } &\makecell[c]{[5] Zhao \\RFIC'25 } \\ \hline
\textbf{Technology} & \textbf{65nm bulk CMOS} & 65nm bulk CMOS & 28nm CMOS & 40nm CMOS & 65nm CMOS \\ \hline
\textbf{Architecture} & \textbf{Three-Tline Series Doherty} & VDPA with Shorted TL & Single TF-Based Parallel Doherty & Balun-first 3-way Parallel Doherty & Balanced PA With ABN \\ \hline
\textbf{Supply (V)} & \textbf{1.0 and 2.0} & 1.2 & 1.8 & 1.1 & 1.2 \\ \hline
\rule{0pt}{0.45cm}\makecell[c]{\textbf{S\(_{21}\)-3dB BW}\\ \textbf{(GHz)}} & \makecell[c]{\textbf{22-32.5}\\ \textbf{(38.5\%)}} & \makecell[c]{21.1-30.4\\(36.1\%)} & \makecell[c]{25.2-29.5\\(15.5\%)} & \makecell[c]{24-30\\(22.2\%)} & \makecell[c]{16.4-22.1\\(29.6\%)} \\ \hline
\textbf{Core area (mm\(^2\))} & \textbf{0.36} & 0.21 & 0.16 & 0.77 & 0.78 \\ \hline
\textbf{Frequency (GHz)} & \textbf{24} & 27 & 27 & 26 & 19 \\ \hline
\textbf{Gain (dB)} & \textbf{22.8} & 19.4 & 16.5 & 20 & 40 \\ \hline
\textbf{P\(_{\text{sat}}\) (dBm)} & \textbf{21.6} & 20.0 & 18.8 & 20.7 & 18.1 \\ \hline
\textbf{PAE\(_{\text{max}}\) (\%)} & \textcolor{red}{\textbf{39}} & 24.6 & 30.1 & 22.3 & 28.8 \\ \hline
\textbf{PAE@6dB PBO} & \textcolor{red}{\textbf{24}} & 20.0 & 22 & 11.7 & 11.2 \\ \hline
\textbf{Modulation scheme} & \textbf{64-QAM} & 64-QAM & 64-QAM & 64-QAM & 64-QAM \\ \hline
\textbf{Data rate (Sym./s)} & \textbf{400M} & 100M & 800M & 400M & 600M \\ \hline
\textbf{EVM (dB)} & \textbf{-26.1} & -25 & -25.0 & -24.9 & -26 \\ \hline
\textbf{P\(_{\text{avg}}\) (dBm)} & \textbf{13.0} & 11.5 & 11.4 & 9.4 & 12.3 \\ \hline
\textbf{PAE\(_{\text{avg}}\) (\%)} & \textbf{17.0} & 14.1 & 18.1 & 14\tnote{*} & 10.2 \\ \hline

\end{tabular}
\begin{tablenotes}
\item[*] Drain efficiency
\end{tablenotes}
\end{threeparttable}
}
\end{table}
\vspace{-0.08cm}
\section*{Acknowledgment}
\vspace{-0.08cm}
This work was supported by the National Natural Science Foundation of China under Grants 62371296, 62188102, and Okawa Foundation Research Grant.
\ifCLASSOPTIONcaptionsoff
  \newpage
\fi



\end{document}